\documentclass[aps,pre,reprint,groupedaddress,longbibliography,floatfix]{revtex4-2}

\usepackage{graphicx} 
\usepackage{dcolumn}
\usepackage{amsmath,amssymb}
\usepackage{physics}
\usepackage[dvipsnames]{xcolor}
\usepackage{times, mathptmx}
\usepackage{bm}
\usepackage[T1]{fontenc}
\usepackage{placeins}
\usepackage{changepage}
\usepackage{nameref}
\usepackage[colorlinks=true, linkcolor=black, citecolor=black, urlcolor=black]{hyperref}

\usepackage{lmodern}
\usepackage{xcolor}
\usepackage{etoolbox}
\usepackage{changes}
\setaddedmarkup{\textcolor{black}{#1}}
\setdeletedmarkup{}

\newcommand{\muh}{$\mu_{\mathrm o}$}
\newcommand{\mua}{$\mu_1$}
\newcommand{\fa}{$\mathrm{F_1}$}
\newcommand{\fh}{${\mathrm{F_{o}}}$}

\newcommand{\na}{$n_1$}
\newcommand{\nh}{$n_{\mathrm o}$}
\newcommand{\pa}{$\mathcal{P}_1$}
\newcommand{\ph}{$\mathcal{P}_{\mathrm o}$}

\begin{document}

\title{Effects of Symmetry on Coupled Rotary Molecular Motors}

\author{Sara Iranbakhsh}
\email{sara\_iranbakhsh@sfu.ca}
\author{David A.\ Sivak}
\email{dsivak@sfu.ca}
\affiliation{Department of Physics, Simon Fraser University, Burnaby, British Columbia V5A 1S6, Canada}

\date{\today}

\begin{abstract}
As engineering advances toward the nanoscale, understanding design principles for molecular motors becomes increasingly valuable. Many molecular motors consist of coupled components transducing one free-energy source into another. Here, we study the performance of coupled rotary molecular motors with different rotational symmetries under constant and scaling driving forces. Under constant driving and strong coupling, symmetry match between the motors decreases the output power. In contrast, under a scaling driving force, the output power is not sensitive to symmetries. However, driving the upstream motor too strongly reduces the downstream motor’s output power, leading to a perhaps counterintuitive phenomenon we term disruption, in which the two motors become disconnected. Across both driving schemes, output power peaks at intermediate coupling, confirming the value of flexible coupling. Beyond providing insights into biological motors, these findings could inform the future design of synthetic nanomotors and structure-based drugs. 
\end{abstract}

\maketitle

As Schrödinger famously articulated, death is the inevitable decay into thermodynamic equilibrium~\cite{schrodingerWhatLifeMind1992}. The nonequilibrium conditions required to sustain life are primarily established and maintained by molecular motors. These nanometer-sized protein complexes perform a wide range of essential functions, including transporting cargo across the cell~\cite{hirokawaKinesinSuperfamilyMotor2009,reck-petersonCytoplasmicDyneinTransport2018,sellersMyosinsDiverseSuperfamily2000}, pulling biopolymers~\cite{raoBacteriophageDNAPackaging2008,singletonStructureMechanismHelicases2007}, enabling muscle contraction~\cite{albertsMolecularBiologyCell2017}, and actively transporting small molecules across membranes~\cite{albertsMolecularBiologyCell2017,jorgensenStructureMechanismNaKATPase2003}.

These multi-component motors operate at low Reynolds number (corresponding to overdamped motion) and receive stochastic kicks from their environment, making directed motion challenging. However, microscopic biological systems have evolved over billions of years to overcome these obstacles, with experimental studies finding efficiencies as high as 90\%~\cite{silversteinExplorationHowThermodynamic2014}. A deeper understanding of these highly efficient biological systems can provide a strong foundation for the design of synthetic molecular motors, as studies show that human-made molecular motors could become significantly more efficient by incorporating bio-inspired mechanisms~\cite{zhangMolecularMachinesBioinspired2018}.

\begin{figure}[t]
    \centering
    \includegraphics[width=0.9\columnwidth]{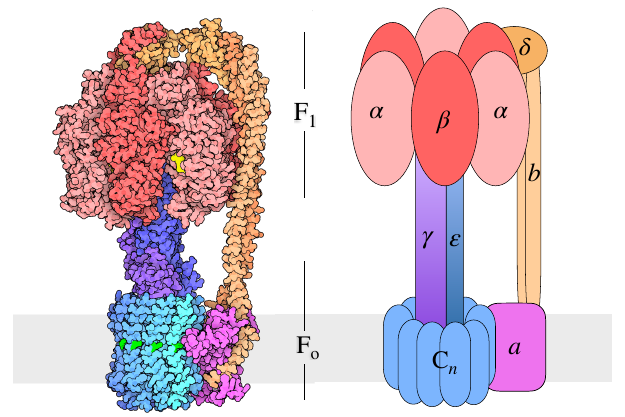}
    \caption{Bacterial ATP synthase structure.  Left: molecular structure, adapted from~\cite{pdb101bioenergy}; right: schematic. $\rm{H}^{+}$ crosses the membrane through \fh, turning the central crankshaft ($\gamma$ and $\varepsilon$). Turning of the central crankshaft causes a conformational change in \fa\ that results in the chemical reaction of ADP and ${\rm P}_{\rm i}$ to produce ATP. \fh\ is mainly the multi-component C-ring C$_n$, which is composed of $n$ subunits. \fa\ has three $\alpha\beta$ subunits across all species.}
    \label{fig:ATP_mine}
\end{figure}

The vast majority of living organisms depend specifically on rotary molecular motors at the cellular level. Well-known examples include the \fh\ and \fa\ motors, which couple to form F-type ATP synthase---responsible for generating the majority of ATP used by the cell~\cite{yoshidaATPSynthaseMarvellous2001}. F-type ATP synthase is the only class of ATPases found in every domain of life~\cite{nirodyATPSynthaseEvolution2020}. Figure~\ref{fig:ATP_mine} shows a schematic of the bacterial ATP-synthase structure. Across all species, the membrane-extrinsic \fa\ component has a three-fold symmetric structure of $\alpha\beta$ subunits, with each subunit creating one ATP per rotation. The rotation of the membrane-embedded \fh\ motor is caused by ions crossing the membrane through the C-ring, driving the central crankshaft around, which in turn rotates within the \fa, catalyzing ATP production. The number of ions crossing the membrane in one full rotation of the \fh\ motor is directly linked to the number of subunits on the C-ring of \fh, with each subunit binding and transporting one ion.

\fh\ and \fa\ display distinct rotational symmetries that directly reflect their subunit composition. The \fa\ motor has a three-fold symmetry, corresponding to its three catalytic subunits. In contrast, the \fh\ motor exhibits a variable symmetry depending on the organism, with the number of C-ring subunits ranging from eight (e.g., in human mitochondria) to seventeen (in bacterium B.\ pseudomallei)~\cite{laiStructureHumanATP2023,pogoryelovC15RingSpirulina2005,kuhlbrandtStructureMechanismsFType2019a,schulzMolecularArchitectureNtype2017}. The symmetry mismatch between \fh\ and \fa, along with the fixed number of subunits in \fa\ and the variable number of subunits in \fh, raise fundamental questions about how evolution may tune ATP synthase to improve its function under varying conditions. While previous studies have proposed methods for modeling interactions between coupled rotary motors~\cite{lathouwersNonequilibriumEnergyTransduction2020}, the role of symmetry mismatch remains largely unexplored.

Symmetry mismatch has been proposed to have functionality for the performance of large protein complexes~\cite{goodsellStructuralSymmetryProtein2000}, including rotary motors. It was hypothesized that symmetry mismatch in bacteriophage structure could facilitate DNA injection~\cite{hendrixSymmetryMismatchDNA1978}. Furthermore, symmetry mismatch is seen in the components of the bacterial flagellar motor (BFM)~\cite{nakamuraStructureDynamicsBacterial2024}, where it is proposed to enhance flexibility and adaptability during assembly~\cite{kawamotoNativeFlagellarMS2021,vartanianStructureFlagellarMotor2012}. 

\FloatBarrier

In this work, we investigate how symmetry match and mismatch influence motor dynamics by incorporating two driving‑force schemes: one that is constant and one that scales with the degree of rotational symmetry. We show that for a constant driving force combined with tight coupling between the motors, the output power shows significant decreases at symmetry match. For a scaling driving force, symmetry-dependent variations were not observed; however, the competition between coupling strength and the driving force became apparent: a strong driving force breaks the communication between the motors, making the tuning of the driving force critical. In both schemes, the output power peaks at intermediate coupling~\cite{lathouwersNonequilibriumEnergyTransduction2020}, pointing to the importance of flexibility in coupling rotary motors, no matter which driving scheme is used.

Our approach provides insights into the trade-offs associated with increasing the number of subunits that supply energy, shedding light on possible design strategies shaped by evolution. A deeper understanding of these mechanisms could inform the design of synthetic rotary molecular motors~\cite{pogoryelovEngineeringRotorRing2012} (with application in medicine, engineering, and beyond~\cite{erbas-cakmakArtificialMolecularMachines2015}), guide the structure-based design of drugs~\cite{andriesDiarylquinolineDrugActive2005,preissStructureMycobacterialATP2015}, improve photosynthetic efficiency in plants~\cite{yamamotoImpactEngineeringATP2023,yiStructureRegulationSignificance2024}, and help study the disorders associated with ATP synthase defects~\cite{dautantATPSynthaseDiseases2018,kuhlbrandtStructureMechanismsFType2019a}.

\emph{Models}.---Denoting the two coupled molecular motors by \fh\ and \fa\ (while this study is motivated by rotary motors \fh\ and \fa\ in ATP synthase, the model is general), the joint probability distribution evolves according to the Fokker-Planck equation,
\begin{align}
\label{eq:FPE} 
    \frac{\partial P(\theta_{o},\theta_{1},t)}{\partial t} &=  
    \frac{1}{\gamma_{\rm o}}
    \left[\frac{\partial}{\partial \theta_{o}}\left( \frac{\partial V}{\partial \theta_{o}} - \mu_{\rm o} \right) 
    +\frac{1}{\beta}\frac{\partial^2}{\partial \theta_{o}^2}\right]P(\theta_{o},\theta_{1},t)  \nonumber \\
    &\hspace{-3em}+ \frac{1}{\gamma_1}
    \left[\frac{\partial}{\partial \theta_{1}}\left( \frac{\partial V}{\partial \theta_{1}} - \mu_1 \right) 
    +\frac{1}{\beta}\frac{\partial^2}{\partial \theta_{1}^2}\right] P(\theta_{o},\theta_{1},t) \ ,
\end{align}
for angular orientations $\theta_{\rm o}$ and $\theta_1$, driving forces $\mu_{\rm o}$ and $\mu_1$ (with opposite signs), and friction coefficients $\gamma_{\rm o}$ and $\gamma_1$ of \fh\ and \fa, respectively. $V$ is the effective potential landscape, which sums the underlying potential landscape of each motor and the coupling potential. \added{Here and henceforth, $\beta =(k_{\rm B}T)^{-1}$, where $k_{\rm B}$ is the Boltzmann constant and $T$ is the temperature.}

Each motor, on its own, can be thought of as a Brownian particle diffusing on an energy landscape with barriers. This energy landscape is most simply modeled as proportional to $\cos n\theta$, where $n$ is the number of barriers. For ATP synthase, jumping over a barrier corresponds to synthesizing or hydrolyzing an ATP molecule (in \fa) or translocation of a proton across the membrane (in \fh).

The coupling potential $V_{\rm c}(\theta_{\rm o}, \theta_1)$ is a function of the motors' angular orientations. A reasonable choice is a spring-like coupling $V_{\rm c}(\theta_{\rm o},\theta_1) \propto \cos(\theta_{\rm o} -\theta_1)$ that favors the in-sync movement of the two motors, giving an effective potential landscape
\begin{align}
    V(\theta_{\rm o},\theta_1) &= -\tfrac{1}{2}E_{\rm o}\cos(n_{\rm o}\theta_{\rm o}) -\tfrac{1}{2}E_1\cos(n_1\theta_1) \notag \\
    &\quad -\tfrac{1}{2}E_{\rm c}\cos( \theta_{\rm o}-\theta_1)\ ,
    \label{energy_land_eq}
\end{align}
for respective barrier numbers \nh\ and \na\ and barrier heights $E_{\rm o}$ and $E_1$ on \fh\ and \fa, and coupling strength $E_{\rm c}$. For ATP synthase, the coupling strength most likely represents the elasticity of the domain at the interface of \fh\ and \fa\ motors, as other domains are rather stiff in comparison~\cite{sielaffDomainComplianceElastic2008,okunoStiffnessSubunitF1ATPase2010,jungeATPSynthase2015}. \added{Note that our model does not explicitly treat the number of bound ions or ion–ion interactions in an ion-driven system like ATP synthase; however, these factors implicitly influence the barrier heights.}

We numerically simulate~\eqref{eq:FPE} using a finite-difference algorithm in time and space (our code is publicly available on Github~\cite{githubver2}, building on~\cite{lathouwersNonequilibriumEnergyTransduction2020,githubEmma}). At steady state, we calculate the average flux $\langle J_{\rm o} \rangle$ and $\langle J_1\rangle$ of \fh\ and \fa, respectively (SM \textup{I}~\cite{SM}). Using the average fluxes, the average input power $\mathcal{P}_{\rm o}$ and the average output power $\mathcal{P}_1$ are~\cite{lathouwersInternalEnergyInformation2022,ehrichEnergyInformationFlows2023} 
\begin{subequations}
\begin{align}
    \mathcal{P}_{\rm o} &= 2\pi \mu_{\rm o} \langle J_{\rm o} \rangle  \label{eq:input} \\
    \mathcal{P}_{1} &= -2\pi \mu_1 \langle J_{1} \rangle \ . 
    \label{eq:output}
\end{align}
\end{subequations}
Moreover, the average slippage flux is defined as the difference between the average fluxes,
\begin{equation}
    J_{\rm slip} \equiv \langle J_{\rm o} \rangle - \langle J_1\rangle \ .
    \label{eq:slippage_flux}
\end{equation}
For convenience, hereafter we omit any explicit mention of averages in describing ensemble-average quantities.

There are different possible choices for coupling the driving forces to the motors, in particular for how the driving forces vary with the number of subunits on a given motor (here we focus on variation of the number of \fh\ subunits). In the BFM, over a wide range of rotational speeds, the torque exerted by the motor on the filament is approximately constant~\cite{xingTorqueSpeedRelationship2006,bergRotaryMotorBacterial2003}; in ATP synthase, the number of ions passing through the membrane in one rotation (and hence the torque supplied to the motor) is proportional to the number of subunits on the C-ring~\cite{wattBioenergeticCostMaking2010,duser36degStepSize2009}. Motivated by such biophysical examples, we consider two types of driving forces: one constant, corresponding to a fixed free-energy change per full rotation, and one that scales with the number of subunits,\added{
\begin{equation}
    \mu_{\rm o/1} = n_{\rm o/1} \times \mu_{\rm o/1}^{\rm sub},
\end{equation}
}corresponding to a fixed free-energy change per subunit \added{$\mu_{\rm o/1}^{\rm sub}$}. \added{For ATP synthase, the driving forces represent the energy provided by ions crossing the membrane in $2\pi$ rotation of \fh\ (\muh), and the cost of creating three ATPs (\mua).} Comparing the results of the constant-force model with the scaling-force model helps in understanding the impact of symmetry on the input power and output power.

\FloatBarrier
\begin{figure*}[htbp]
    \centering
    \includegraphics[width=\textwidth]{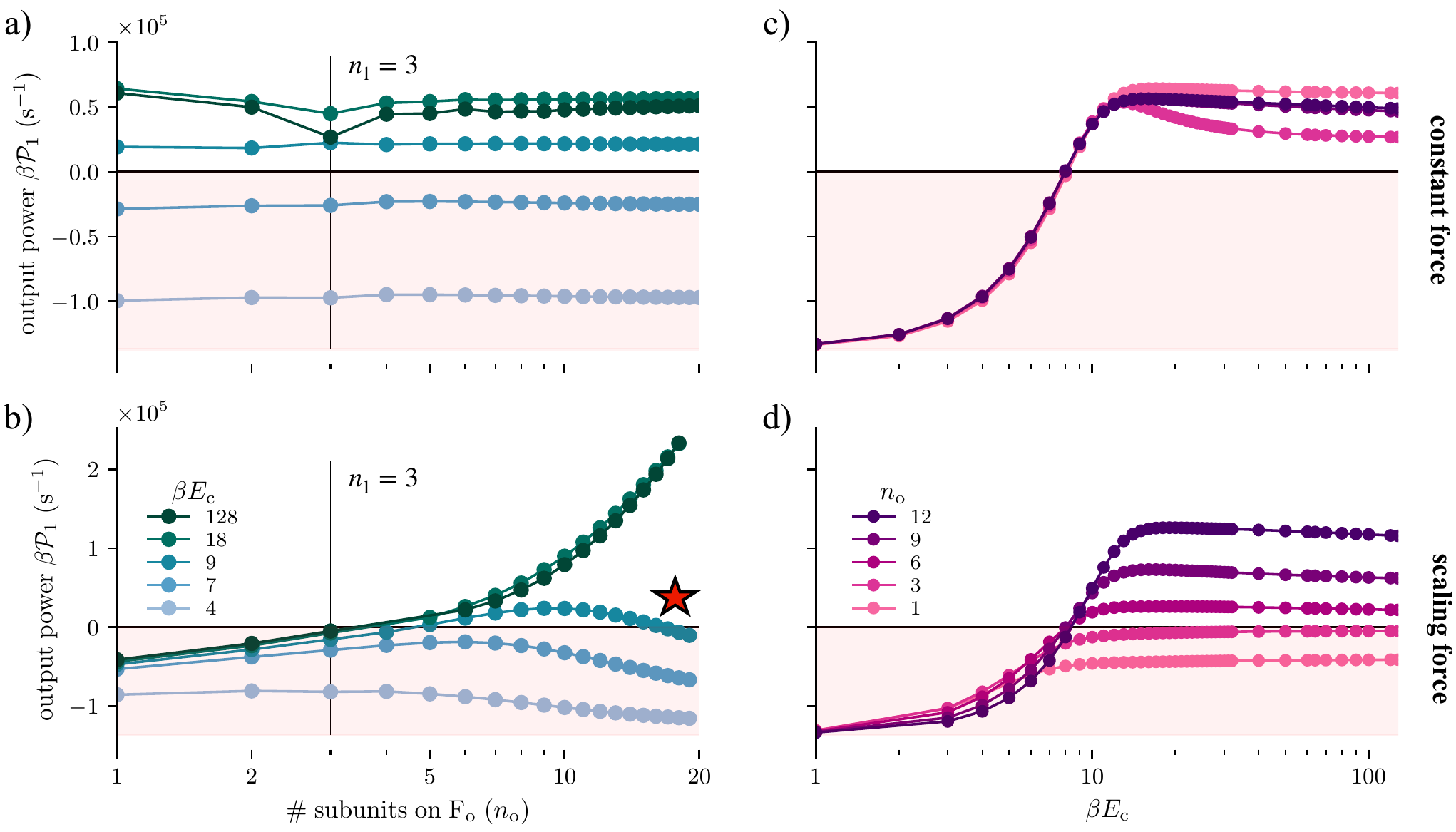}
    \caption{a,b) Output power as a function of the number \nh\ of \fh\ subunits, for different coupling strengths $\beta E_{\rm c}$. Vertical lines indicate $n_1 = n_{\rm o} = 3$. Star labels disruption. c,d) Output power as a function of the coupling strength, for different numbers \nh\ of \fh\ subunits. a,c) Constant driving forces $\beta\mu_{\rm o} = 4$ and $\beta \mu_1 = -2$. b,d) Scaling driving forces $\beta \mu_{\rm o} = 0.5 \times n_{\rm o}$ and $\beta \mu_1 = -2/3\times 3 = -2$. $n_1 =3$ \added{and $\beta E_{\rm o} = \beta E_1 = 2$} throughout. \added{Here and from now on, shared axes have the same scales.}}
    \label{fig:P1_vs_EC_vs_n0}
\end{figure*}

\emph{Varying the number of \texorpdfstring{\fh}{Fo} subunits}.---For a single rotary motor on a periodic landscape, under a constant driving force, as the number of subunits (and hence barriers) increases, the flux slightly decreases and then saturates at a many-subunit limit; however, under a scaling driving force, flux increases linearly with subunit number (SM~\textup{II}~\cite{SM}). These trends suggest that the magnitude of the driving force has a greater influence on the flux than the motor's internal structure (number of barriers or metastable states) does (SM~\textup{III}~\cite{SM}). 

A similar effect is seen in the case of coupled rotary motors. Under a constant driving force, the number of subunits on the upstream motor \fh\ (specified in the constant-force scheme, without loss of generality, by $\mu_{\rm o} > -{\mu_1}$) does not generally affect the output power (Fig.~\ref{fig:P1_vs_EC_vs_n0}a): \fa\ is mostly insensitive to the number of subunits of \fh\ except when $n_{\rm o}$ is an integer multiple of $n_1$ and most significantly at symmetry match ($n_{\rm o}= n_1$). At symmetry match, it is harder for both \fa\ and \fh\ to cross their coinciding barriers, giving a lower flux $J_1$ and hence lower output power when the two motors are sufficiently strongly coupled. In fact, for tight coupling, the lowest output power occurs where the symmetries are matched (the numbers of subunits are equal). For intermediate coupling ($\beta E_{\rm c}\approx 10$), the output power is slightly higher for equal numbers of subunits than for differing numbers (excluding the single subunit case), since the greater flexibility allows \fh\ to jump forward first and help \fa\ subsequently jump. In this intermediate-coupling regime, the symmetry match is beneficial, although the increase in the output power is relatively small (SM~\textup{IV}~\cite{SM}).

Figure~\ref{fig:P1_vs_EC_vs_n0}b shows that a scaling force produces distinctive behavior compared to a constant force. Using this scheme for applying the driving forces, the number of subunits of both motors becomes rather irrelevant compared to the effect of increasing the driving force of \fh. Starting from weak driving forces (small $n_{\rm o}$) for any non-zero $E_{\rm c}$ value, the flux $J_1$ increases as a larger driving force speeds the rotation of the downstream motor, so long as the coupling to the upstream motor is sufficiently strong.

\emph{Disruption}.---The coupling $E_{\rm c}$ between the two motors competes with the driving force, which tends to decouple them. Under a scaling driving, for a fixed coupling strength, increasing the number of subunits on \fh\ (and hence achieving a larger \muh) does not always increase the output power, as the fixed coupling becomes less restraining for both motors due to the increase in the driving force. In fact, for each given $E_{\rm c}$, there is a critical number of subunits beyond which output power decreases and eventually becomes negative. 

We define \emph{disruption} as this change from positive to negative output power while maintaining positive input power: a disrupted coupled rotary system is one in which energy is consumed by both \fh\ (positive \ph) and \fa\ (negative \pa), failing to perform useful work. For a given $E_{\rm c}$, disruption occurs because the driving force on \fh\ becomes so large that the fixed coupling no longer dominates over slippage. SM~\textup{V}~\cite{SM} provides an analytical treatment of disruption in coupled barrierless motors.

\begin{figure}[t]
    \centering
    \includegraphics[width=\columnwidth]{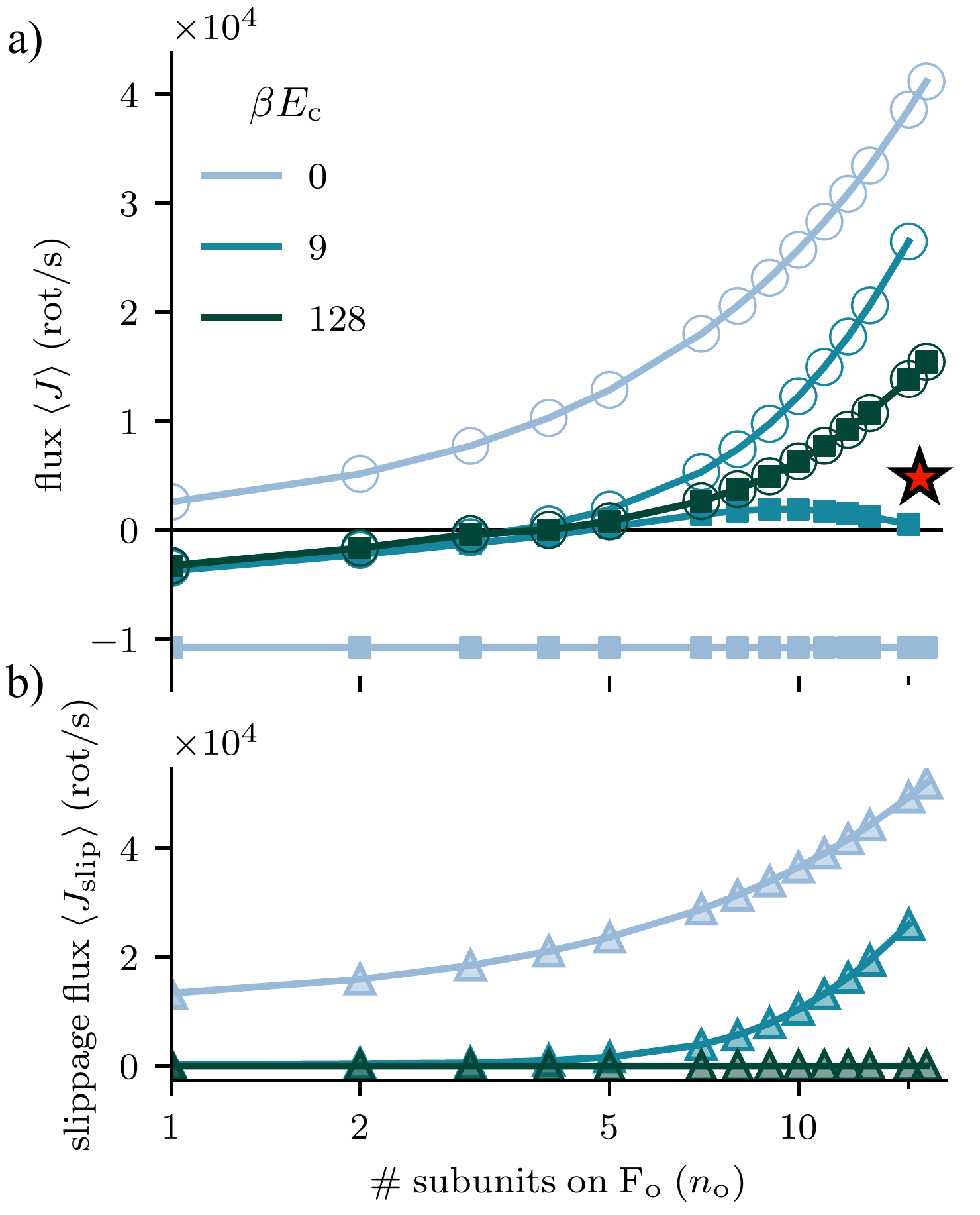}
    \caption{Flux as a function of the number \nh\ of \fh\ subunits, for different coupling strengths $\beta E_{\rm c}$, using a scaling driving force. a) Hollow circles: \fh\ flux $J_{\rm o}$; solid squares: \fa\ flux $J_1$. Star labels disruption. b) Slippage flux $J_{\rm slip}$. $n_1 =3$\added{, $\beta \mu_{\rm o} = 0.5 \times n_{\rm o}$, $\beta \mu_1 = -2/3 \times 3 = -2$}\added{, and $\beta E_{\rm o} = \beta E_1 = 2$} throughout.}
    \label{fig:J1_J0_Jslip}
\end{figure}

Figure~\ref{fig:J1_J0_Jslip} shows how disruption emerges from a significant increase in the slippage flux~\eqref{eq:slippage_flux}. Due to the large \muh, \fh\ moves fast, so \fa\ spends almost the same amount of time being pulled (through coupling to \fh) in each direction. Therefore, the two motors are practically disconnected: each rotates in the direction of its respective individual driving force. 

\begin{figure}[htbp]
    \centering
    \includegraphics[width=\columnwidth]{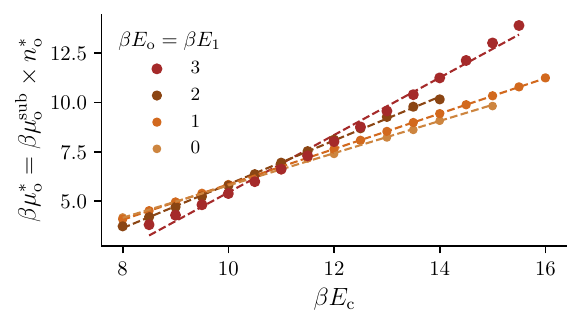}
    \caption{The scaling driving force $\beta \mu_{\rm o}^{*}$ that maximizes output power, as a function of coupling strength $\beta E_{\rm c}$, for various barrier heights. Dashed lines are linear fits. \deleted{$\beta \mu_1 = -2$ and} $n_1 = 3$\added{, $\beta \mu_{\rm o} = 0.5 \times n_{\rm o}$, and $\beta \mu_1 = -2/3 \times 3 = -2$} throughout.}
    \label{fig:mu_star_vs_Ec}
\end{figure}

Figure~\ref{fig:P1_vs_EC_vs_n0}b points to the existence, for each coupling strength, of a maximum power at some intermediate $n_{\rm o}$ that we label $n_{\rm o}^{*}$. Figure~\ref{fig:mu_star_vs_Ec} shows that the power-maximizing driving force ($\mu_{\rm o}^{*}$) is empirically roughly linear in the coupling strength, across different barrier heights. 

\begin{figure}[t]
    \centering 
    \includegraphics[width=\linewidth]{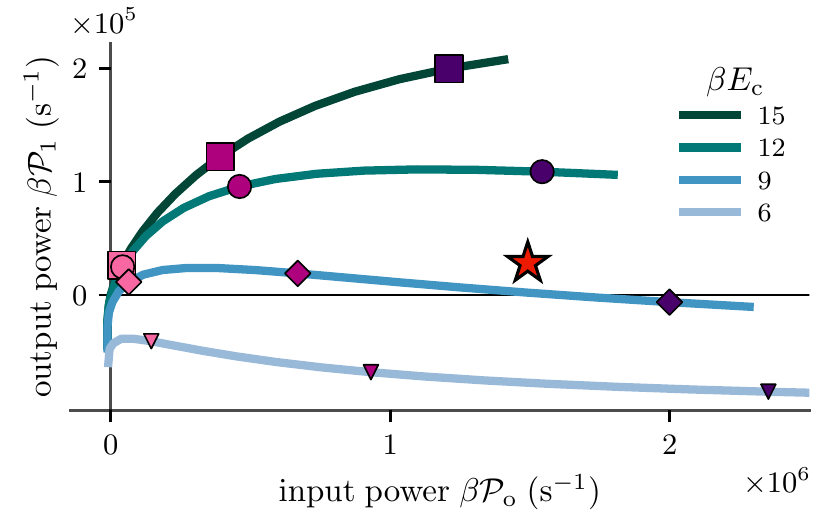}
    \caption{Parametric plot of output power and input power with varying $n_{\rm o}$ under a scaling driving force \muh, for several coupling strengths $\beta E_{\rm c}$. As a guide to the eye, curves connect points for different \nh. Points with shades of purple represent $n_{\rm o} = 6,\ 12,\ 18$ from light to dark. Star labels disruption. \deleted{$\beta \mu_1 = -2$ and }$n_1 = 3$\added{, $\beta \mu_{\rm o} = 0.5 \times n_{\rm o}$, $\beta \mu_1 = -2/3 \times 3 = -2$}\added{, and $\beta E_{\rm o} = \beta E_1 = 2$} throughout.} 
    \label{fig:out_v_in}
\end{figure}

For a scaling driving force, Fig.~\ref{fig:out_v_in} shows how output power and input power vary parametrically with $n_{\rm o}$, for different coupling strengths. For a fixed range of \fh\ subunits (and thus \muh), higher coupling strength ensures a monotonic input-output relation, whereas weaker coupling produces nonmonotonic behavior. Thus, under intermediate coupling, fewer subunits on the upstream motor can yield higher output power. 

This aspect is particularly important in the design of coupled rotary motors. Previous work suggests that the coupling strength between \fh\ and \fa\ in ATP synthase is independent of the ring size (\nh)~\cite{pogoryelovEngineeringRotorRing2012}. Consequently, in designing synthetic coupled rotary motors---where tuning the coupling strength may not be feasible---understanding how \pa\ varies with \ph\ at a specific coupling strength becomes crucial for evaluating whether increasing the input is justified.

\emph{Varying the coupling strength}.---The behavior of output power as the coupling strength $\beta E_{\rm c}$ changes shows notable features when comparing the two models. In the weak-coupling limit, as a significant connection has not been established between the two motors, each motor rotates in the direction of its own driving force, and the composite system can be considered as having two isolated rotary motors. On the other hand, in the tight-coupling limit, both motors move in lockstep in the direction of the largest driving force.  But, while the behavior of the coupled rotary motor can be analytically studied for tight coupling and no coupling, the behavior for intermediate coupling depends on many factors, such as the barrier height, the ratio of driving forces, and symmetry. 

Under a constant force, having the least number of subunits (and hence barriers) is beneficial; hence, the output power is maximized for $n_{\rm o} = 1$. For all curves seen in Fig.~\ref{fig:P1_vs_EC_vs_n0}c, power is maximized at intermediate coupling strength as intermediate coupling gives flexibility in coordinating the angular orientations of the two motors~\cite{lathouwersNonequilibriumEnergyTransduction2020}, which allows the motors to benefit from independent thermal kicks while maintaining energy transduction (see SM~\textup{VI}~\cite{SM} for a detailed discussion on symmetry effects in the constant-driving-force scheme). 
The sharp peak in the $n_o = n_1 =3$ curve in Fig.~\ref{fig:P1_vs_EC_vs_n0}c shows that a composite system of two motors with the same number of subunits is most sensitive to variation of coupling strength. This is because a symmetry match produces a constructive landscape overlap that requires more careful tuning of the coupling to ensure \fh\ leads \fa\ without significant slippage (e.g., \fh\ rotating without \fa\ following).

On the other hand, under a scaling driving force, there is no sensitivity to the number of subunits; the output power only significantly depends on the driving force. In this scheme (as for a constant force), the output power is maximized at intermediate coupling strength (Fig.~\ref{fig:P1_vs_EC_vs_n0}d).

\emph{Conclusion and Discussion}.---In this paper, we studied the effects of symmetry mismatch on the performance of coupled rotary motors, incorporating two biologically motivated driving-force schemes: one that is constant and one that scales with the number of subunits. We focused     on output power, as it is a key energetic consideration when designing a system of coupled rotary motors, as illustrated by natural examples. For instance, ATP molecules fuel the cell's processes~\cite{neupaneATPSynthaseStructure2019}, making fast ATP turnover essential. Therefore, understanding how to optimize output power becomes crucial. Other important performance metrics include the efficiency (SM~\textup{VII}~\cite{SM}).

Under the constant-driving scheme, we found that at tight coupling, symmetry match is disadvantageous to output power, resulting in considerably slower rotation in the downstream motor. This finding is valuable in designing coupled motors operating in environments with fixed available free energy: for coupled rotary motors with inherently tight coupling between mechanical, chemical, or other degrees of freedom, symmetry match reduces operational speed and should therefore be avoided.

For a scaling driving force, we did not observe any symmetry-dependent variation in the rotational speed of \fa, as the performance change with varying parameters, such as coupling strength, was the same for symmetry match as for symmetry mismatch. However, with a scaling driving force, the competing effects of driving and coupling were apparent, showing disruption in the coupling between the motors. Thus, it becomes important to tune the driving to ensure that the motors remain coupled while maximizing the rotational speed of the downstream motor.

In both of the driving schemes, for any number of subunits on the upstream motor, the output power is maximized at intermediate couplings, as seen in~\cite{lathouwersNonequilibriumEnergyTransduction2020}. In case of a symmetry match, the rotational speed of the downstream motor is more sensitive to changes in the coupling strength when using a constant driving force.

Our findings contribute to a growing effort to understand coupled nonequilibrium stochastic systems~\cite{sloweySloppyGearMechanism2022, jimenez-pazInformationThermodynamicsCellular2025}, and provide a framework for optimizing coupled rotary motors with distinct symmetries. These findings can be applied in designing synthetic molecular motors and structure-based drugs, enhancing photosynthetic efficiency, and studying the evolution of molecular-motor structures. Furthermore, the principles explained here have applications in studying other coupled molecular motors. For example, the BFM exhibits similar dynamics, rotating its rotor by moving ions through multiple stators. The connection between the stators (upstream motors) and the BFM’s rotor (downstream motor) can be analyzed in a similar manner using nonequilibrium statistical approaches, facilitating the quantitative analysis of complex biological systems.

\vspace{2ex}
We thank Antonio Patrón Castro (SFU Physics) and Matthew Leighton (Yale Physics) for their insights and comments on the manuscript. This work was supported by an SFU Graduate Dean's Entrance Scholarship (SI), a Natural Sciences and Engineering Research Council of Canada (NSERC) Discovery Grant and Discovery Accelerator Supplement RGPIN-2020-04950, an NSERC Alliance International Collaboration Grant ALLRP-2023-585940, and a Tier-II Canada Research Chair CRC-2020-00098 (all DAS). This research was enabled in part by support provided by BC DRI Group and the Digital Research Alliance of Canada (www.alliancecan.ca).

\vspace{2ex}
The data that support the findings of this article are openly available at~\cite{iranbakhsh_2025_17316573}.

\FloatBarrier
\nocite{reimannGiantAccelerationFree2001}
\nocite{lathouwersEnergyInformationFlows2021b}
\bibliography{Thesis}

\clearpage
\onecolumngrid

\begin{center}
	\textbf{\large Supplemental Material for ``Effects of Symmetry on Coupled Rotary Molecular Motors''}
\end{center}
\setcounter{equation}{0}
\setcounter{figure}{0}
\setcounter{table}{0}
\setcounter{page}{1}
\makeatletter
\renewcommand{\theequation}{S\arabic{equation}}
\renewcommand{\thefigure}{S\arabic{figure}}


\section{Local and average probability flux}
\label{app:flux}

At the steady-state distribution $P_{\rm st}(\theta_{\rm o},\theta_1)$, the local probability flux of motors \fh\ and \fa\ is 
\begin{subequations}
\label{eq:prob_flux}
\begin{align}
J_{\rm o}(\theta_{\rm o}, \theta_1) &= \frac{1}{\gamma_{\rm o}} \left[\left(\mu_{\rm o}-\frac{\partial V}{\partial \theta_{\rm o}}\right)P_{\rm st}(\theta_{\rm o},\theta_1) - \frac{1}{\beta}\frac{\partial P_{\rm st}(\theta_{\rm o},\theta_1)}{\partial \theta_{\rm o}}\right] \label{eq:prob_fux_0} \\
J_1(\theta_{\rm o}, \theta_1) &= \frac{1}{\gamma_{1}} \left[\left(\mu_1-\frac{\partial V}{\partial \theta_1}\right)P_{\rm st}(\theta_{\rm o},\theta_1) - \frac{1}{\beta}\frac{\partial P_{\rm st}(\theta_{\rm o},\theta_1)}{\partial \theta_1}\right] \ .\label{eq:prob_fux_1}
\end{align}
\end{subequations}
Using the local fluxes, the motors' average fluxes are
\begin{subequations}
\begin{align}
\langle J_{\rm o} \rangle &= \frac{1}{2\pi} \int_0^{2\pi} \int_0^{2\pi} J_{\rm o}(\theta_{\rm o}, \theta_1) \, \dd\theta_1 \, \dd\theta_{\rm o} \\
\langle J_1 \rangle &= \frac{1}{2\pi} \int_0^{2\pi} \int_0^{2\pi} J_1(\theta_{\rm o}, \theta_1) \, \dd\theta_{\rm o} \, \dd\theta_1 \ .
\end{align}
\end{subequations}
The average fluxes are used to calculate input power (3a), output power (3b), and slippage flux (4).

\section{Single rotary motor on a tilted landscape}
\label{app:isolated}
\begin{figure}[h]
    \centering
    \includegraphics[width=0.6\columnwidth]{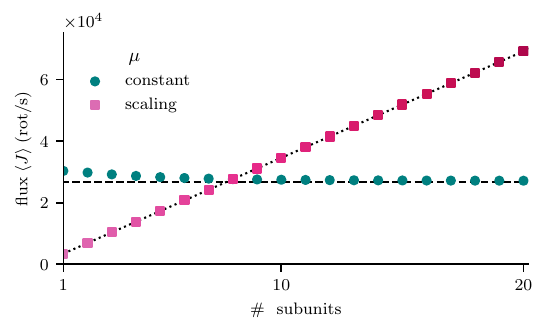}
    \caption{Flux of a single isolated rotary motor as a function of the number of subunits (barriers), for barrier height $\beta E = 1$. Teal points: constant driving force $\beta \mu = 4$. Pink points: scaling driving force $\beta \mu = 0.5 \times n$. Dashed line: asymptotic limit, for constant driving force $\beta\mu = 4$~\eqref{eq:asymptotic_speed}. Dotted line: Analytical flux for scaling driving force $\beta \mu = 0.5 \times n$~\eqref{eq:green_analytical}.} 
    \label{fig:isolated}
\end{figure}

The dynamics of an isolated individual rotary motor provides initial insight into the role of symmetry (Fig.~\ref{fig:isolated}). Rotational speed in a periodic potential $V = -\tfrac{1}{2}E\cos n\theta$ (with barrier height $E$) that is tilted has been analyzed in~\cite{reimannGiantAccelerationFree2001} and introduced in the following form:
\begin{subequations}
\begin{align}
    \langle \dot{\theta} \rangle &= \frac{1 - e^{-\frac{2\pi}{n} \beta \mu}}{{\frac{n}{2\pi}}\int_{0}^{2\pi/n} \dd\theta \, I(\theta)} \label{eq:speed}  \\
    I(\theta) &\equiv \beta \gamma\int_{\rm 0}^{2\pi/n} \dd\theta' \exp \left\{ \beta \left[  V(\theta) - V(\theta - \theta') - \theta' \mu \right] \right\} \ , 
\end{align}
\end{subequations}
where $D = \frac{1}{\beta \gamma}$. Using the speed, power is $\mathcal{P} =\mu \langle \dot{\theta} \rangle$. 

Numerical simulations of~\eqref{eq:speed} show that, as the number of subunits increases under a constant driving force, the power slightly decreases to a constant in the many-subunit limit (SM~\ref{app:asymptotic}). However, under a scaling driving force, the power increases linearly (SM~\ref{app:green_line}). The difference between the two schemes arises because, although increasing the number of barriers reduces flux---since the motor encounters more obstacles and, on average, takes longer to complete each rotation---this effect is negligible compared to the positive impact of a larger driving force. The insignificance of the number of subunits compared to the driving force is further reinforced by how output power in coupled motors changes with \fh's driving force in the constant- and scaling-driving schemes (SM~\ref{app:P1_vs_force}).

\subsection{Analysis of speed in the constant-driving-force scheme with many subunits}
\label{app:asymptotic}
Here we evaluate the asymptotic behavior of \eqref{eq:speed} in the limit of a large number $n$ of barriers, under the constant-driving-force scheme. 

The change of variables
\begin{subequations}
\begin{align}
    \phi &= n \theta\\
    \quad \phi' &= n \theta'\ ,
\end{align}
\end{subequations}
maps the integration domain to \(\phi, \phi' \in [0, 2\pi]\). Taylor expanding~\eqref{eq:speed} in the small parameter $\frac{1}{n}$ gives
\begin{align}
\langle \dot{\theta} \rangle 
&\approx \left[ \frac{4\pi^2 \mu}{\gamma} \left(1 - \frac{\pi \beta \mu}{n}\right) \right] \times \notag \\
&\quad \left\{ 
   \int_0^{2\pi} \int_0^{2\pi} A(\phi, \phi') \, \dd\phi' \, \dd\phi \right. 
 \left. 
   - \frac{\beta \mu}{n} \int_0^{2\pi} \int_0^{2\pi} \phi' A(\phi, \phi') \, \dd\phi' \, \dd\phi \right. 
 \left.
   + \frac{(\beta \mu)^2}{2n^2} \int_0^{2\pi} \int_0^{2\pi} {\phi'}^2 A(\phi, \phi') \, \dd\phi' \, \dd\phi 
\right\}^{-1} \ ,
\label{eq:pink_analytical}
\end{align}
for the force-independent base integrand
\begin{equation}
A(\phi, \phi') \equiv \exp\left\{ 
\tfrac{1}{2} \beta E [\cos(\phi - \phi') - \cos \phi] \right\} \ .
\end{equation}

Under constant driving, as the number $n$ of barriers increases the average rotational speed saturates to the asymptotic value
\begin{equation}
\langle \dot{\theta} \rangle_{\infty} = \frac{4\pi^2 D \beta \mu}{\displaystyle\int_0^{2\pi} \int_0^{2\pi} A(\phi, \phi') \, \dd\phi' \, \dd\phi} \ .
\label{eq:asymptotic_speed}
\end{equation}

\subsection{Analysis of speed in the scaling-driving-force scheme}
\label{app:green_line}
For a scaling driving force ($\mu = n \mu^{\text{sub}}$), \eqref{eq:speed} 
becomes
\begin{align}
\langle \dot{\theta} \rangle &= \frac{2\pi \left( 1 - e^{-2\pi \beta \mu^{\text{sub}}} \right)}{\beta \gamma\int_0^{2\pi} \int_0^{2\pi} 
e^{- \beta \mu^{\text{sub}} \phi'} \, A(\phi, \phi')
\, \dd\phi' \, \dd\phi} \times n \ .
\label{eq:green_analytical}
\end{align}
Thus, under a scaling driving force, the average rotational speed grows linearly with $n$.

\section{Effects of the number of subunits of \texorpdfstring{\fh}{Fo} and its driving force on the output power}
\label{app:P1_vs_force}

\begin{figure}[t]
    \centering  \includegraphics[width=0.6\columnwidth]{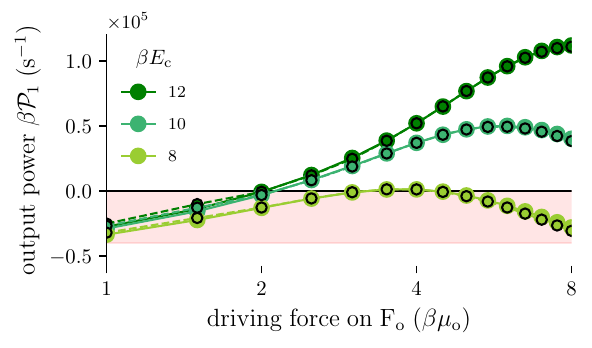}
    \caption{Output power as a function of the \fh\ driving force $\mu_{\rm o}$, for different coupling strengths $\beta E_{\rm c}$. Solid lines with solid markers: constant driving force with $n_{\rm o} =8$. Dashed lines with black-edged markers: scaling driving force ($\beta\mu_{\rm o}^{\rm sub} = 0.5$). $n_1 =3$\added{, $\beta E_{\rm o} = \beta E_1 = 2$,} and $\beta\mu_1=-2$ throughout. }
    \label{fig:P1_vs_force}
\end{figure}

The number of subunits of the motors, in comparison to the driving forces, appears to be of little importance. Figure~\ref{fig:P1_vs_force} shows the output power for both constant- and scaling-driving-force schemes. In the constant-force scheme, the number of subunits on \fh\ and \fa\ is kept fixed while the magnitude of the driving force is increased. The figure shows striking similarity between this and the scaling-force scheme---where the number of subunits on \fh\ increases alongside the driving force. 

This finding suggests that architectural differences influence output power predominantly through their effect on the driving force. When the driving force is held fixed, varying the number of subunits generally has little impact on output power.

\section{Symmetry match at constant driving for intermediate and tight coupling}
\label{app:symmetry_match_constat_drive}

Under a constant driving force, the coupling strength significantly influences output power, as the interference of the individual energy landscapes is constructive at symmetry match (where $n_{\rm o} = n_1$). 

\begin{figure}[t]
    \centering
    \includegraphics[width=0.6\columnwidth]{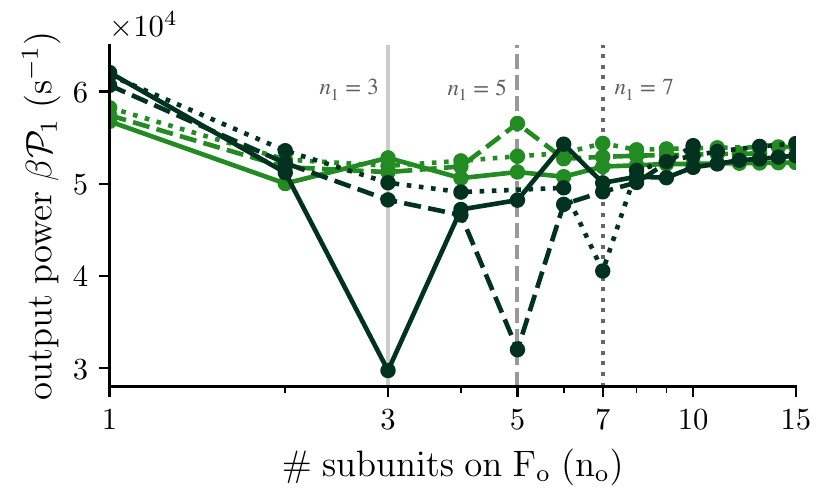}
    \caption{Output power as a function of \nh\ for constant driving force. $n_1 =$ 3 (solid), 5 (dashed), and 7 (dotted). Light green points: $\beta E_{\rm c} = 12$. Dark green points: $\beta E_{\rm c} = 50$. \added{$\beta E_{\rm o} = \beta E_1 = 2$, }$\beta \mu_{\rm o} =4$ and $\beta \mu_{1} =-2$ throughout. Vertical lines from light to dark: $n_1 =$ 3, 5, 7.}
    \label{fig:symmetry_match_at_3_5_7}
\end{figure}

Figure~\ref{fig:symmetry_match_at_3_5_7} illustrates output power as a function of the number of \fh\ subunits. At symmetry match, the output power exhibits distinct behavior for different coupling strengths: intermediate coupling yields a local maximum, while tight coupling results in a global minimum. These findings underscore the critical role of symmetry match in modulating output power (SM~\ref{app:symm_const}).

\section{Analytical investigation of disruption in barrierless rotary motors} 
\label{app:analytical_disruption}

To keep the treatment in this paper self-contained, this section summarizes the derivations explained in detail in \cite{lathouwersEnergyInformationFlows2021b}. For coupled barrierless rotary motors (i.e., with flat individual potentials), mathematical manipulations of the individual energy landscapes lead to an effective energy landscape for half the separation $\Delta \theta \equiv \tfrac{1}{2}(\theta_{\rm o} - \theta_1)$ of angular orientations:
\begin{equation}
    V_{\text{eff}}(\Delta\theta) \equiv -\tfrac{1}{4} E_{\text{c}} \cos 2\Delta \theta - \mu \Delta \theta \ .
\end{equation}
Here $\mu \equiv \tfrac{1}{2} (\mu_{\rm{o}} - \mu_1)$.
Fluxes can be analytically calculated:
\begin{subequations}
\begin{align}
    J_{\rm o} &= {J}_{\bar\theta} +  {J}_{\Delta \theta} \label{eq:analytical_Barrirless_J0} \\ 
    J_{\rm 1} &= {J}_{\bar\theta} -  {J}_{\Delta \theta} \label{eq:analytical_Barrirless_J1}\ ,
\end{align}
\end{subequations}
where
\begin{subequations}
\begin{align}
{J}_{\bar\theta} &= \frac{\mu_{\mathrm{o}} + \mu_{\mathrm{1}}}{4\pi \gamma}\label{eq:J_bar_theta} \\
{J}_{\Delta \theta} &= \frac{k_{\mathrm{B}} T}{2\gamma} \left\{ \vphantom{\int_0^{2\pi}} \right. 
\left( 1 - e^{-2\pi \beta \mu} \right)^{-1} \int_0^{2\pi} \dd\theta \, e^{\beta V_{\mathrm{eff}}(\theta)} 
\int_0^{2\pi} \dd\theta' \, e^{-\beta V_{\mathrm{eff}}(\theta')} 
\left. - \int_0^{2\pi} \dd\theta \, e^{-\beta V_{\mathrm{eff}}(\theta)} 
\int_0^{\theta} \dd\theta' \, e^{\beta V_{\mathrm{eff}}(\theta')}
\right\}^{-1} \label{eq:J_delta_theta} \ .
\end{align}
\end{subequations}
with $\bar \theta \equiv \tfrac{1}{2}(\theta_{\rm o} + \theta_1)$.

\begin{figure}[t]
    \centering
    \includegraphics[width=0.65\columnwidth]{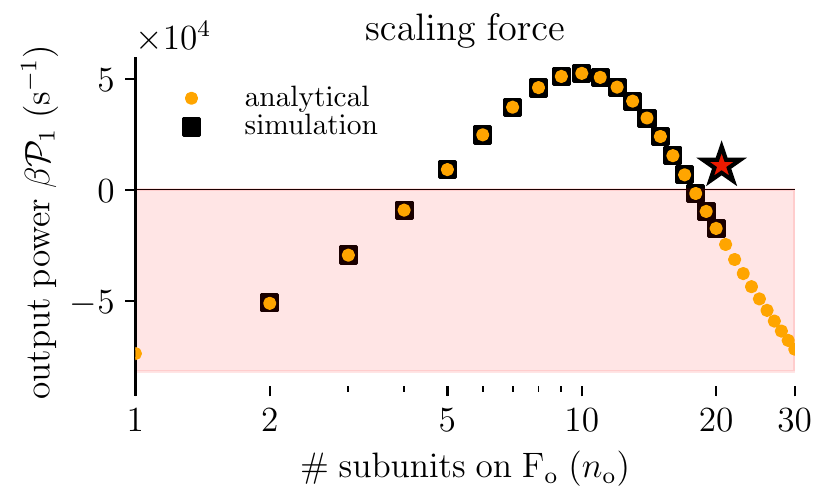}
    \caption{Disruption in barrierless coupled motors under a scaling-driving-force scheme. Squares: analytic results using \eqref{eq:analytical_Barrirless_J1}; circles: simulations. Star labels disruption. Increasing the \fh\ subunit number amplifies the driving force, leading to disruption at any coupling ($\beta E_{\rm c} = 9$ shown). \added{$\beta \mu_{\rm o} = 0.5 \times n_{\rm o}$, $\beta \mu_1 = -2/3 \times 3 = -2$, and} \added{$\beta E_{\rm o} = \beta E_1 = 2$ throughout.}}
    \label{fig:barrierless-analytical-sim}
\end{figure}

Figure~\ref{fig:barrierless-analytical-sim} compares for $\beta E_{\rm c} = 9$ the simulated output power with the analytical solutions, but any other coupling strength would show the overlap of the analytical and simulation results for a barrierless system. Apart from providing a test of consistency, Fig.~\ref{fig:barrierless-analytical-sim} demonstrates disruption in energy transduction without energy barriers, showing how disruption is a result of an excessive driving force on the upstream motor and not the interference of the individual motors' energy landscapes.

\section{Symmetry in constant-driving-force scheme}
\label{app:symm_const}

\subsection{The power-maximizing coupling strength}
\label{app:power_max_EC}
\begin{figure}[t]
    \centering
    \includegraphics[width=0.6\columnwidth]{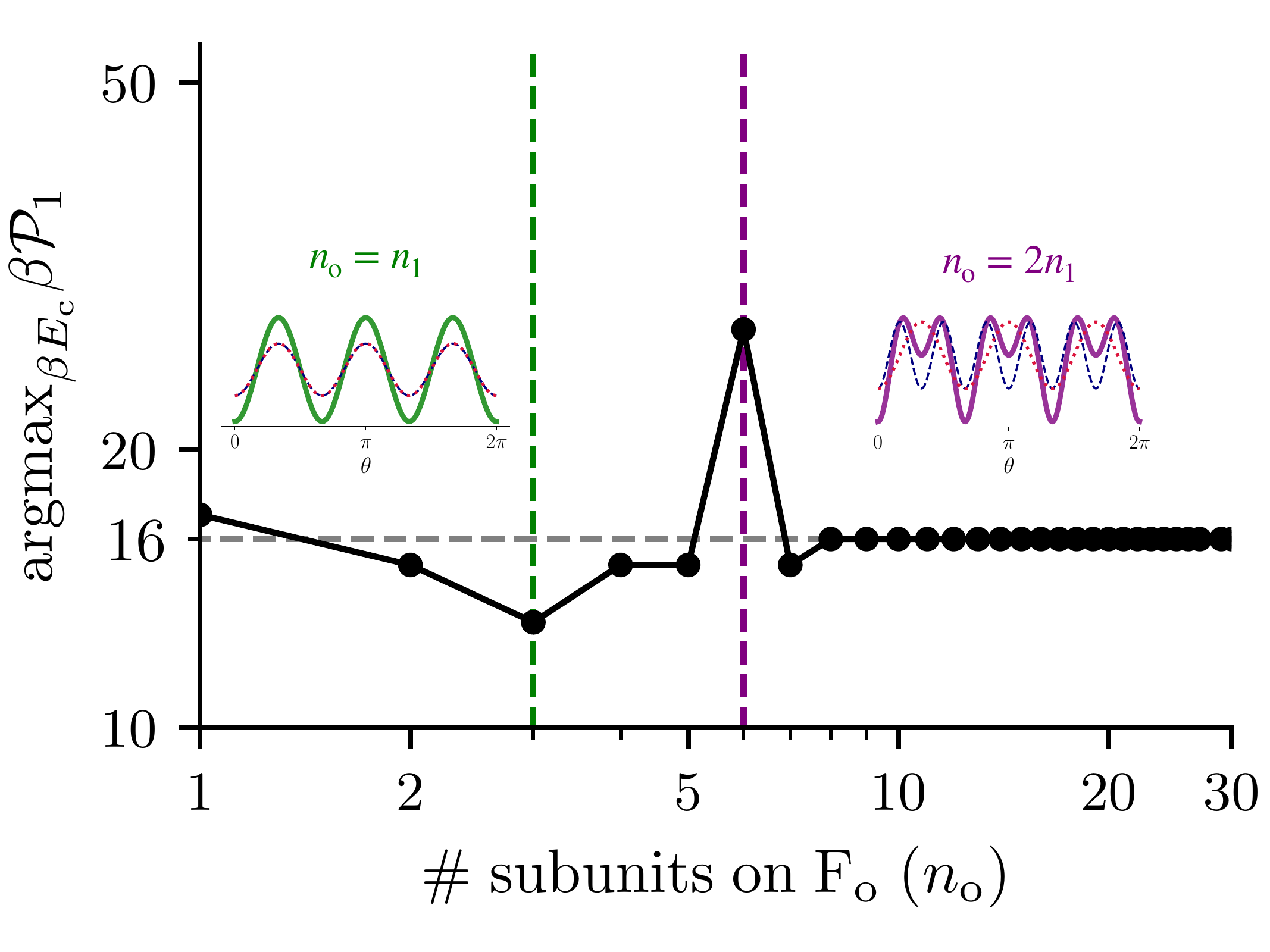} 
    \caption{The power-maximizing coupling strength as a function of \nh\ under a constant driving force. Output power peaks at an intermediate coupling for all \nh, mostly at $\beta E_{\rm c} = 16$ (gray dashed line). Green and purple landscapes indicate the sum of the individual landscapes of \fh\ (blue and dashed) and \fa\ (red and dotted) at symmetry match ($n_{\rm o} = n_1$) and the first integer multiple of $n_1$ ($n_{\rm o} = 2 n_1$). \added{$\beta E_{\rm o} = \beta E_1 = 2$ throughout.}} 
    \label{fig:power_maximizing_Ec}
\end{figure}

Figure 2c demonstrates how at any \nh\, there is an intermediate coupling strength that maximizes the output power. Figure~\ref{fig:power_maximizing_Ec} shows these power maximizing coupling strengths as a function of the subunits on \fh. For all values of \nh, output power is maximized at intermediate coupling strength, with greater \nh\ having the same power-maximizing coupling strength. But, when \nh\ is comparable to \na, the symmetry mismatch between the two motors affects the optimum coupling.

Equal subunits on the two motors result in a constructive potential landscape that makes jumping over the barriers harder. Hence, having a symmetry match requires more freedom in coupling for moving out of the potential wells, which explains the dip at $n_{\rm o} = n_1 =3$ seen in Fig.~\ref{fig:power_maximizing_Ec}. Unequal numbers of subunits (e.g, $n_{\rm o} = 6, \ n_1 =3$) lead to a misalignment between the energy landscapes of the two motors, which allows \fh\ excessive freedom of movement, during which \fa\ is not always pulled along. This excessive freedom can be compensated for by using a tighter grip (coupling strength), hence the sharp peak seen in Fig.~\ref{fig:power_maximizing_Ec}.

SM~\ref{app:inch_slip} provides an analytical approach to explaining why the coupling that maximizes the output power is stronger at symmetry mismatch compared to symmetry match.

\subsection{Dominant-paths analysis}

\label{app:inch_slip}
An analytical approach to understanding the effect of symmetry on performance is to examine the dominant pathways available to the motors. The method adopted here is inspired by~\cite{lathouwersNonequilibriumEnergyTransduction2020}, which identified two primary pathways governing the collective dynamics of the motors at symmetry match ($n_1 = n_{\rm o} = 3$): inchworming and full slippage.

\begin{figure}[htbp]
    \centering
    \includegraphics[width=0.65\columnwidth]{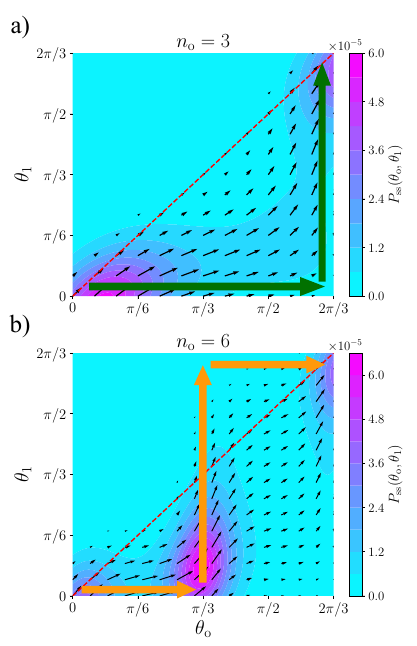}
    \caption{Probability flux~\eqref{eq:prob_flux} (black arrows) and steady-state probability (heatmap) across one barrier ($0$ to $2\pi/3$) of the downstream motor with fixed $n_1=3$. a) At symmetry match ($n_{\rm o} = 3$), the most likely events are full slippage (\fh\ completing a full rotation without driving \fa) and inchworming (following the green arrows). b) At symmetry mismatch ($n_{\rm o} = 6$), additional pathways such as hop-catch (following the orange arrows) become probable. \added{$\beta E_{\rm o} = \beta E_1 = 2$, }$\beta\mu_{\rm o} =4$, $\beta\mu_1 =-2$, $n_1 =3$, and $\beta E_{\rm c} = 12$ throughout.}
    \label{fig:prob_quiver}
\end{figure}

\begin{figure}[t]
    \centering 
    \includegraphics[width=0.7\columnwidth]{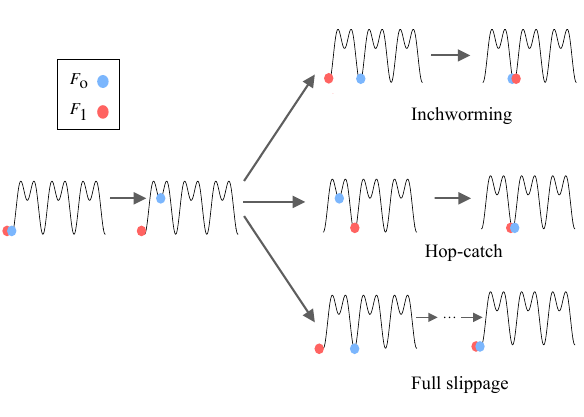}
    \caption{The most likely pathways for a system with 6 subunits on \fh\ and 3 subunits on \fa. The effective potential landscape (excluding torques) is depicted schematically, with \fh\ and \fa\ represented by blue and red circles, respectively. In the \emph{inchworming} pathway, \fh\ first moves ahead by $2\pi/3$, followed by \fa\ inchworming forward. In the \emph{hop-catch} pathway, when \fh\ is $2\pi/6$ ahead of \fa, \fa\ first hops over \fh, and then \fh\ catches up to \fa. In \emph{full slippage}, \fh\ completes one full rotation without any movement from \fa.}
    \label{fig:model1_landscape_paths}
\end{figure}

Figure~\ref{fig:prob_quiver}a shows that at symmetry match and with sufficiently flexible coupling, the upstream motor (\fh) moves forward first and either pulls the downstream motor (\fa) along via the coupling (inchworming) or continues rotating separately, completing its cycle without dragging along the downstream motor (full slippage).

These two event types constitute the dominant dynamical pathways at symmetry match; however, for symmetry mismatch, notably when \nh\ is an integer multiple (greater than one) of \na, a third event also significantly contributes: \emph{hop-catch}. In a hop-catch event, \fh\ transitions to a metastable state that does not align with one of \fa’s potential wells. The coupling then causes \fa\ to hop ahead of \fh, reaching its own metastable state. Finally, \fh\ catches up and joins \fa\ at their now-coinciding metastable states (Fig.~\ref{fig:prob_quiver}b). 

Using these three most likely pathways of inchworming, hop-catch, and full slippage, we follow the trajectory of the two motors, assuming they start from the same rotational angle (Fig.~\ref{fig:model1_landscape_paths}). For simplicity, only two integer multiples of $n_1 = 3$ ($n_o = 3, 6$) are considered.  

The output power is approximated using the respective rates of pathways that affect \pa, inchworming (at rate $r^{\text{inch}}$) and hop-catch ($r^{\text{hc}}$):
\begin{equation}  
    \mathcal{P}_1 \propto \frac{-2\pi\mu_1}{3}(r^{\rm inch} + r^{\rm hc}) \ .  
    \label{N/t}
\end{equation} 
(The slippage pathway does not contribute to \pa.)
\begin{figure}[t!]
    \centering
    \includegraphics[width=0.6\columnwidth]{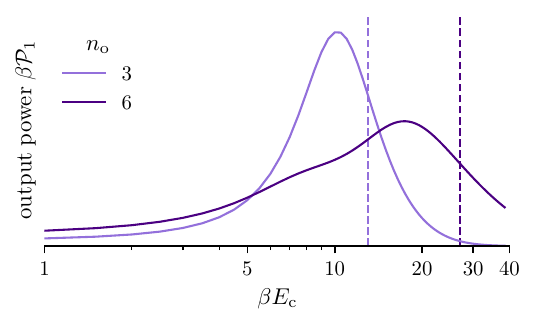}
    \caption{Output power \pa\ as a function of the coupling strength $\beta E_{\rm c}$ using branch analysis for the upstream motor having \nh\ = 3 or 6. Vertical dashed lines: power-maximizing $\beta E_{\rm c}$ from simulations (Fig. 2c). $n_1 =3$ \added{and $\beta E_{\rm o} = \beta E_1 = 2$} throughout.}
    \label{fig:model1_branch_analytical}
\end{figure}
Figure~\ref{fig:model1_branch_analytical} illustrates how the output power changes with $E_{\rm c}$. The power-maximizing coupling strength seems to be dependent on the symmetry ratio $n_{\rm o}/n_1$. This approach correctly predicts significantly higher power-maximizing coupling at $n_{\rm o} = 6$ as seen in Fig.~\ref{fig:power_maximizing_Ec}.

\begin{figure*}[htbp]
    \centering
    \includegraphics[width=\textwidth]{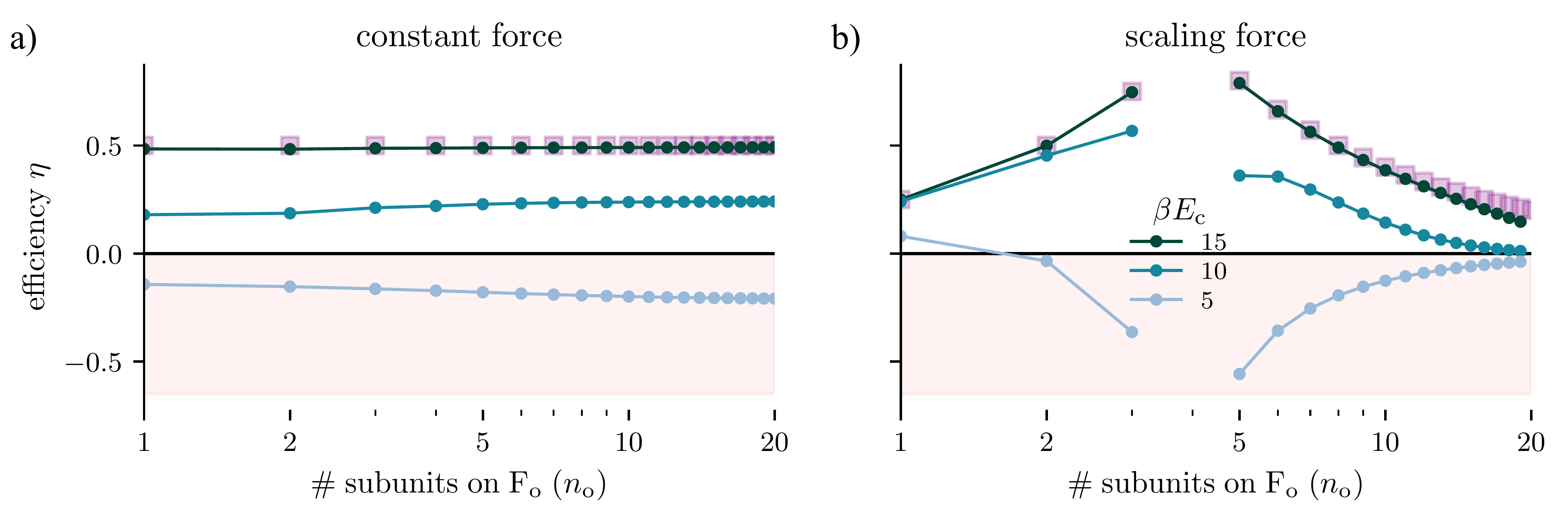}
    \caption{Efficiency as a function of the number \nh\ of \fh\ subunits for different coupling strengths $\beta E_{\rm c}$. a) Constant driving forces $\beta\mu_{\rm o} = 4$ and $\beta \mu_1 = -2$. b) Scaling driving forces $\beta \mu_{\rm o} = 0.5 \times n_{\rm o}$ and $\beta \mu_1 = -2/3\times 3 = -2$. Efficiency is undefined at the stall point ($n_{\rm o} = 4$). Purple squares: $\eta^{\rm max}$. $n_1=3$ \added{and $\beta E_{\rm o} = \beta E_1 = 2$} throughout.}
    \label{fig:efficiency_vs_n0}
\end{figure*}

\section{Efficiency as a function of the number of \texorpdfstring{\fh}{Fo} subunits}
\label{app:efficiency}
Efficiency is an important performance metric in studying coupled rotary molecular motors. Here, we define efficiency as the ratio of the smaller power (that of the downstream motor) to the larger power (that of the upstream motor):
\begin{equation}
    \eta = \frac{\mathcal{P}^{\rm down}}{\mathcal{P}^{\rm up}} \ .
    \label{eq:eta_def}
\end{equation}
Since at very strong coupling strength $E_{\rm c}$ the coupling term in the energy landscape dominates, the result is coordinated movement between the two motors. In this tight-coupling limit, as the two motors move in lockstep, efficiency is maximized at $\eta^{\rm max}$ and can be analytically calculated. \added{Note that in this tight-coupling limit, efficiency is undefined at the stall point (i.e., where the driving forces \muh\ and \mua\ have opposite signs but equal magnitudes), since there is no net flux for either motor.}

For the constant-driving scheme at infinitely tight coupling, and assuming $\mu_{\rm o} > -\mu_1$,
\begin{equation}
    \eta^{\rm max}_{\rm const} = \frac{-\mu_1}{\mu_{\rm o}} \ . \label{eq:eta_tight_const}
\end{equation}
For the scaling driving force, since labeling motors as upstream and downstream is based on the relative strengths of their driving forces, and the driving of \fh\ changes with \nh, efficiency at infinitely tight coupling becomes
\begin{subequations}
\begin{align}
    \eta^{\rm max}_{\rm scaling} &= \frac{-\mu_1}{\mu_{\rm o}} \ , \ \ \ \ \ {\rm if} \ \mu_{\rm o} > -\mu_1  \\
    \eta^{\rm max}_{\rm scaling} &= \frac{\mu_{\rm o}}{-\mu_1} \ , \ \ \ \ \ {\rm otherwise} \ .
\end{align}
\end{subequations}

Figure~\ref{fig:efficiency_vs_n0}a shows how the constant-driving-force efficiency varies with \nh: at a fixed coupling strength, efficiency is not impacted significantly by the number of subunits on the upstream motor. In contrast, Fig.~\ref{fig:efficiency_vs_n0}b shows strong dependence on \nh\ when the driving force on \fh\ is scaled by the number of subunits. As a result, \nh\ influences key factors such as which motor is more strongly driven and whether the chosen coupling strength is sufficiently loose to cause system disruption at larger subunit numbers.
\end{document}